\documentclass{iaus}
\usepackage{graphicx}
\usepackage{epstopdf}

\title[Statistical Tools of ISM turbulence] 
{Statistical Tools of Interstellar Turbulence: Connecting Observations with Theory}

\author{B. Burkhart \& A. Lazarian}  

\affiliation{Astronomy Department, University of Wisconsin, Madison, 475 N. 
Charter St., WI 53711, USA
 \\ email: {\tt burkhart@astro.wisc.edu}}

\pubyear{2010}
\volume{274}  
\pagerange{1--4}
\jname{Advances in Plasma Astrophysics}
\editors{A.Bonanno,  E. de Gouveia Dal Pino \& A. Kosovichev, eds.}
\begin{document}

\maketitle

\begin{abstract}
MHD Turbulence is a critical component of the current paradigms of star formation, particle transport, magnetic reconnection  and evolution of the ISM, to name just a few.
Progress on this difficult subject is made via numerical simulations and observational studies, however in order to connect these two, statistical
methods are required.  This calls for new statistical tools to be developed in order to study
turbulence in the interstellar medium.  Here we briefly review some of the recently developed statistics that focus on characterizing
gas compressibility and magnetization and their uses to interstellar studies. 

\end{abstract}

\section{Introduction}
The paradigm of the interstellar medium has undergone major shifts in the past two decades thanks to the combined efforts of high resolution surveys and the exponential
increase in computation power allowing for more realistic numerical simulations.  The ISM is now known to be
highly turbulent and magnetized, which affects  ISM structure, formation, and evolution.   Magnetohydrodynamic turbulence is essential
to many astrophysical phenomena such as star formation, cosmic ray dispersion,  magnetic reconnection, and many transport processes.
However the study of turbulence is complicated by the fact that  no complete theory for turbulence exists.  

To this date, turbulence has always been understood in a statistical manner - showing the 'order from chaos.'
 The classical picture of Kolmogrov 1941 depicts turbulent flows as composed of eddies which transfer energy across a range of scales, typically from the larger injection
 scale down to what is know as the dissipative range. The  eddies 
 are unstable and break into smaller and smaller  eddies,  transferring their kinetic energy in a manner that conserves it over time.  
 This picture of the 'energy cascade'  is highly dependent on the characteristic
dimensionless Reynolds number, which describes the ratio of the inertial to viscous forces.  By simple dimensional analysis it can be shown that
the energy spectrum scales as $k^{-5/3}$. More current research shows that  turbulence is intermittent and not statistically self-similar, as
is assumed in the Kolmogrov picture.  To add to the complexity, the ISM is in a plasma state and gas dynamics are governed by the MHD equations,
 which changes the the scaling relationships via
interactions of MHD waves and Alfv\'enic shearing.

In addition to the Reynolds number, other important parameters of ISM MHD turbulence include the sonic Mach number, which describes
the compressibility of the gas as ${\cal M}_s \equiv {\bf v}/C_s$, and the Alfv\'enic Mach number ${\cal M}_A\equiv {\bf v}/v_A $, where 
$v_A = |{\bf B}|/\sqrt{\rho}$ is the Alfv\'enic velocity, ${\bf B}$ is magnetic field and $\rho$ is density.  These parameters are not always easy to characterize observationally, with the
Alfv\'enic Mach number being particularly difficult due to cumbersome observational measurements of vector magnetic field.  

Several techniques have been around for decades in order to study ISM
turbulence and its properties. Many of these hinge on either density fluctuations,
via scintillation in ionized media, Radio
position-position-velocity data (PPV) and column density
maps for neutral media.  The advantage of spectroscopic
data is that it contains information about the
turbulent velocity field as well as the density fluctuations, however separation of the 
density and velocity fields is non trivial (Lazarian 2009). When invoking the phenomena of turbulence  
many researchers base their analysis around
 the slope of the log-log spatial power spectrum.  While the power spectrum
gives information about the energy per wavenumber (or frequency), it only contains the Fourier amplitudes 
and completely ignores the phases.  This alone is motivation for the development of complimentary techniques.  

\begin{figure*}[tbh]
\centering
\includegraphics[scale=.4]{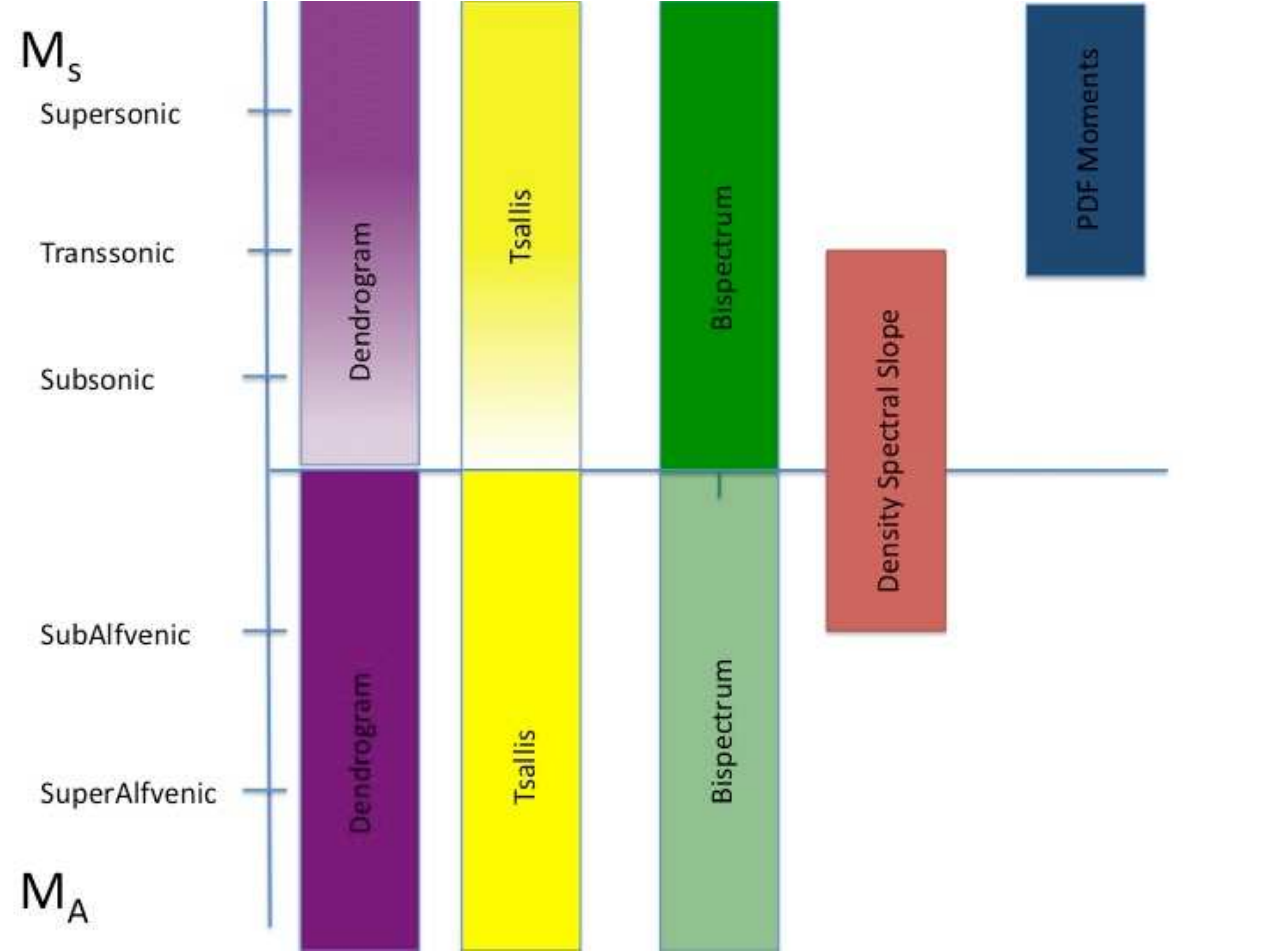}
\caption{Cartoon showing different statistics studied with their dependencies on the sonic and Alfv\'enic Mach numbers.  The different colors indicate
different staitistics.  The intensity of the colors indicate the confidence the statistics can provide for the parameter on the y-axis in our simulations.}
\label{fig:ps}
\end{figure*}
In general, the best strategy for studying a difficult subject like interstellar turbulence is to use a synergetic approach,  combining theoretical knowledge, numerical simulations,
and observational data via statistical studies.  In this way one can obtain the most complete and reliable picture of the physics of turbulence.  
Here we seek to extend the statistical comparison between numerical and observational turbulence by reviewing statistical tools that can
greatly compliment the information provided by the power spectrum.   In particular we focus the review on 
tools that can provide information on ${\cal M}_s$ and ${\cal M}_A$.  This review will highlight several different tools (see Figure 1) studied in the works of Kowal et al. 2007, 
Burkhart et al. 2009, 2010,2010b, Esquivel et al. 2010, Toffelmire et al. 2010, Chepurnov \& Lazarian 2009, Rosolowsky et al. 2008, 
 and Goodman et al. 2009,  which represent a mixture of numerical and observational
studies.   We focus on statistics that have application for observable data (PPV or column density).

\section{Supersonic versus Subsonic Turbulence}
The sonic Mach number describes the ratio of the flow velocity to the sound speed, and thus is a measure of the compressibility of the medium.  
Turbulence that is supersonic displays very different characteristics from subsonic turbulence in terms of the spectral slope and density/ velocity 
fluctuations.   Because the physical environment of compressible turbulence is very different from incompressible, this parameter is extremely
important for many different fields of astrophysics including, but not limited to, star formation and cosmic ray acceleration.  
\subsection{Higher Order Moments of Column Density}
Moments of the density distribution can be used to roughly determine the gas compressibility through shock density enhancements.  As the ISM
media transitions from subsonic to supersonic and becomes increasingly supersonic, shocks create enhanced mean value and variance of the
density PDF.  In addition, as the shocks become stronger, the PDF is skewed and becomes more kurtotic than Gaussian.  However, Kowal et al. 2007 showed
that this method is not very effective for subsonic cases, as these distributions are roughly Gaussian.  For areas where Mach numbers approach and 
exceed unity, higher order moments of column density PDFs can be used as a measure of compressibility.

\subsection{Spectrum}
The power spectrum has been used for studies of both observational and numerical turbulence for decades.  In addition to providing information
on the energy cascade, the spectrum can also be used to obtain compressibility in column density maps.  As turbulence transitions from subsonic
to supersonic, density enhancements due to shocks create small scale structures which shallow out the spectral slope.  Additionally, the presence
of a strong magnetic field can create Alfv\'enic shearing, which can steepen the slope via destruction of small scale structures. 
Subsonic low magnetic field spectral slope values are close to what is predicted for hydrodynamic
Kolmogorov turbulence and the addition of a strong magnetic field and shocks cause deviations from the predicted -11/3.  
\footnote{ For incompressible turbulence,
the Kolmogorov power spectrum in three dimensions (3D) is ~$k^{-11/3}$, in 2D 
it is $k^{-8/3}$, and 1D $k^{-5/3}$  for the same energy spectrum E(k).}
Burkhart et al. 2010
compared the spectral slope found for the SMC galaxy with simulations of turbulence and found that the SMC's slope of -3.3  matches well with 
transsonic type turbulence.  This Mach number range has ben independently confirmed by observational method to obtain the sonic number, which
utilize the ratio of the HI spin to kinetic temperature, and also other statistical approaches such as the higher order moments.

\subsection{Bispectrum}
While the power spectrum has been used extensively in ISM studies, higher order spectrum have been more rare.  The bispectrum, or Fourier transform of the
3rd order autocorrelation function, has been applied to isothermal ISM turbulence simulations and the SMC only recently (Burkhart et al. 2009, 2010) although
it is extensively used in other fields including cosmology and biology.

The bispectrum preserves both the amplitude and phase and  provides information on the interaction of wave modes.  Completely randomized modes will
show a bispectrum of zero, while mode coupling will show non-zero bispectrum.   Shocks and high magnetic field have been shown to increase mode coupling in the 
bispectrum.  Due to their ability to shallow out the density energy spectrum, shocks greatly enhance the small scale wave-wave coupling.  
With simulations with the same sonic number, mode correlation is shown to increase  with an increase in magnetic field, however this is less clear for column density
as it is for 3D density.

\section{Magnetization of Turbulence}
The Alfv\'en number is the dimensionless ratio of the flow velocity to the Alfv\'en speed.  As the Alfv\'en speed depends on the magnetic field, this
ratio can provide information on the strength of the magnetic field relative to the velocity and density.  The Alfv\'enic number is critical in several fields including
interplanetary studies and star formation.  The solar wind is known to be a  super-Alfv\'enic flow while the Alfv\'enic number in star forming
regions is still hotly debated. 
\subsection{Tsallis PDFs of PPV and Column Density}
PDFs of increments (of density, magnetic field, velocity etc.) are a classic way to study turbulence since the phenomena is scale dependent. 
The Tsallis function was formulated in Tsallis 1988  as a means to extend traditional Boltzmann-Gibbs mechanics 
to fractal and multifractal systems. 

\begin{equation}
R_{q}= a \left[1+(q-1) \frac{\Delta f(x,r)^2}{w^2} \right]^{-1/(q-1)}
\label{eq:1}
\end{equation}

 The Tsallis distribution  (Equation 3.1) can be fit to PDFs of increments, that is, $f(x,r)=G(x + r) - G(x)$, where G(x) is a particular field (for example, turbulent density, velocity
 or magnetic field).  The Tsallis fit parameters (q, a, and w in Equation 3.1) describe the width, amplitude, and tails of the PDF.  These parameters have been shown
to have dependencies on both sonic and Alfv\'enic Mach number.  Tsallis parameters are able to distinguish between sub, tras and supersonic
turbulence as well as gauge wether the turbulence is sub-Alfv\'enic or super-Alfv\'enic.  The parameter that describes the width of the PDF distribution
is particularly sensitive. 

\subsection{ Dendrograms of Position-Position-Velocity (PPV) data}
A Dendrogram (from the Greek dendron ÒtreeÓ,-
gramma ÒdrawingÓ) is a hierarchical tree diagram that
has been used extensively in other fields, particular
galaxy evolution and biology. It is a graphical representation
of a branching diagram, and for our particular
purposes with PPV data, quantifies how and where local
maxima of emission merge with each other. The dendrogram
was first used on ISM data in Rosolowsky et al.
2008 and Goodman et al. 2009 in order to characterize
self-gravitating structures in star forming molecular
clouds.

Burkhart et al. 2010 used the dendrogram on synthetic PPV cubes and found it to be rather sensitive to magnetic density/velocity enhancements.
They looked at the moments of the distribution of local maxima in emission found in the tree diagram.   These moments showed clear signs
of  being dependent on Mach numbers, with the particular strength being the Alfv\'en number.
When high frequency filtering is applied in order to mask small scale enhancements due to shocks, the magnetic enhancements
fully dominate the moments of the tree diagram distribution.   The dendrogram is also able to distinguish between simulations that
show varying degrees of gravitational strength.  
It is also very encouraging that this
statistic is working in PPV space, while other statistics studied utilize the column density maps. This further motivates the synergetic approach of using these
statistics.

\section{Conclusions}
The last decade has seen major increases in the knowledge of the ISM and of its turbulent nature thanks to high resolution observations and advanced numerical simulations.  
This calls for new advances in statistical tools in order to best utilize the wealth of observational data in light of numerical and theoretical predictions. 
Recently several authors have explored new tools for studying turbulence beyond the power spectrum.   While these proceedings do not cover all the useful
tools in the literature, we attempt to provide some review on tools that describe the gas compressibility and the Alfv\'enic Mach number by utilizing density fluctuations
created by shocks and magnetic density enhancements.


\begin{thebibliography}{}
\bibitem[Burkhart et al.(2009)]{burkhart09}
Burkhart, B., Falceta-Goncalves, D., Kowal, G., Lazarian, A., 2009, ApJ, 693, 250
\bibitem[Burkhart et al.(2010a)]{burkhart10}
Burkhart, B., Stanimirovic, S., Lazarian, A., Kowal, G., 2010, ApJ, 708, 1204
\bibitem[Burkhart et al.(2010b)]{burkhart10}
Burkhart, B., Goodman, A., Lazarian, A., Rosolowsky, E., 2010, in prep.
\bibitem[Chepurnov(2008)]{}
Chepurnov, A., Lazarian, A., Gordon, J., \& Stanimirovic., S., 2008,
ApJ, 688, 1021
\bibitem[Esquivel(2010)]{es10}
Esquivel, A., \& Lazarian, A., 2010, ApJ, 710, 125
\bibitem[Goodman et al.(2009)]{good09}
Goodman, A. A., Rosolowsky, E. W., Borkin, M. A., Foster, J. B., Halle, M.,
Kauffmann, J. \& Pineda, J. E.,  2009, Nature letters, 457, 63
\bibitem[Lazarian(2006)]{laz06}
Lazarian, A. 2009, Space Science Reviews, 143, 357
\bibitem[Kowal2010]{kowal10}
Kowal, G., Lazarian, A. \& Beresnyak, A., 2007,ApJ, 658, 423
\bibitem[Toffelmire et al.(2010)]{tof10}
Toffelmire B., Burkhart, B., Lazarian, A., 2010, submitted.
\bibitem[rosolossky(2008)]{ros2008}
Rosolowsky, E. W., Pineda, J. E., Kauffmann, J., \& Goodman,
A. A. 2008, ApJ, 679, 1338
\end{thebibliography}
\end{document}